\documentclass[12pt,preprint]{aastex}
\usepackage{graphicx}
\begin{document}
\title{Evolution of Spectral States of Aql X-1 during the 2000 Outburst}

\author{Dipankar Maitra and Charles D. Bailyn}
\affil{Yale University, Department of Astronomy, P.O.Box 208101, New Haven, CT, 06520-8101}
\email{maitra,bailyn@astro.yale.edu}

\begin{abstract}

We present the results of detailed analysis of X-ray data in 3-20 keV range
from a $\sim70$ day outburst of the neutron star transient Aquila X-1 
during September-November 2000. Optical monitoring with the YALO 
1m telescope was used to trigger X-ray observations with {\it Rossi X-ray 
Timing Explorer} (RXTE) in order to follow the outburst from a very early 
stage. In this paper we discuss the correlated evolution in time of features 
in the spectral and temporal domains, for the entire outburst. The state 
transition from low/hard state to high/soft state during the rise of the 
outburst occurs at higher luminosity than the transition back to low/hard 
state during the decay, as has also been observed in other outbursts. Fourier 
power spectra at low frequencies show a broken power law continuum during the 
rising phase, with the break frequency increasing with time. During the 
decline from maximum the source evolves to a position in the hardness-intensity
 plane as well as in the color-color diagram which is similar to, but 
distinct from, the canonical high/soft state. High frequency quasi-periodic 
oscillations from 636-870 Hz were seen only during this transitional state.

\end{abstract}

\keywords{accretion, accretion disks --- stars: neutron --- 
X-rays: binaries --- individual (Aquila X-1)}

\section{Introduction}

Soft X-ray transients (SXTs) are an important class of low-mass X-ray binaries 
(LMXBs). A characteristic property of SXTs is that their luminosity is 
generally below the detection limit of most X-ray telescopes. However 
occasionally they brighten up by several orders of magnitude at all 
wavelengths. Usually X-ray radiation during the low luminosity, low $\dot M$ 
state shows a powerlaw spectrum, suggesting a nonthermal origin of the 
radiation. For SXTs with a neutron star (NS) as the accretor, sometimes a soft 
thermal blackbody component can also be seen which is attributed to surface 
emission from the NS \citep{cs2000,r1999,r2000}. The origin of the powerlaw is 
generally attributed to softer photons that have been upscattered by their 
interaction with hot electrons in a Comptonizing corona (e.g. 
\citet{st1980, n1995, n2002}; however see \citet{mff2001} who suggest that the 
powerlaw photons are due to synchrotron emission from jets). The spectra at 
higher luminosities and hence at higher $\dot M$, however, are largely thermal 
and can be well fit by assuming a disk shaped blackbody with a variable radial 
temperature profile given by $T\sim R^{-3/4}$ \citep{ss1973,m1984}. The 
dominance of hard powerlaw photons in the low luminosity state gives it the 
name {\it low/hard} (LH) state, whereas dominance of soft thermal photons in 
the high luminosity state gives it the name {\it high/soft} (HS) state. The LH 
and HS states are the two canonical spectral states of most SXTs. There are a
host of other spectral states observed which are various combinations of the 
powerlaw and thermal spectra, differing in their relative strengths as well as 
spectral shape (e.g. change in powerlaw photon index $\Gamma$). While 
$\dot M$ is likely one of the factors causing spectral state 
transitions, recent work \citep{h2001,shs2002,mc2003} suggests the existence of 
factor(s) other than $\dot M$. Thus we are faced with a second parameter 
problem in explaining the spectra of SXTs and other X-ray binaries.
 
Aql X-1 has the shortest recurrence time of any SXT known so far, which makes 
it an excellent system to study. Aql X-1 has been observed to go into outburst 
approximately once every year, but the periodicity is not fixed \citep{s2002}. 
Type I X-ray bursts from Aql X-1 \citep{koy1981,ccg1987} suggests that the 
compact object is a weakly magnetized neutron star. The companion is a Roche 
lobe filling late K star \citep{cilp1999,wry2000} and the system has an 
orbital period of 18.95 hrs \citep{wry2000}, at a distance of 4-6.5 kpc 
\citep{r2001}. Based on its evolution on the color-color plane \citep{hk1989}, 
and its spectral and temporal properties, Aql X-1 has been classified as an 
`atoll' source \citep{rmkf2000}.

In September 2000, Aql X-1 initiated a new outburst. We obtained PCA data 
triggered by optical observations. The RXTE data for this outburst have been 
analyzed by \citet{y2003}, who reported hard X-ray flares preceding the 
transition to the soft state and 1-20 Hz QPO in the outburst rise. From the 
similarity in spectral/temporal behaviour between this outburst of Aql X-1 and 
other black hole transients, \citet{y2003} suggest a similar origin for QPOs 
and the continuum shape of PDS for both neutron star transients as well as 
black hole transients.

Here we present our results of spectral and temporal analysis of the data 
obtained during the 2000 outburst of Aql X-1. We find that initially the 
spectrum was dominated by hard powerlaw photons with $\Gamma < 2$, with little 
or no disk. As the source luminosity increased, the source made a very rapid 
transition from LH state to a HS state, dominated by the disk and also a steep
($\Gamma > 3$) falling powerlaw. However, during the decaying phase, the 
transition back to LH state occurred at a lower luminosity (and hence lower 
inferred mass accretion rate) than the transition from LH to HS state. This 
phenomenon has been observed for other outbursts of Aql X-1, as well as other 
SXTs (\citet{m1995,mc2003} and references therein). Thus the source traces a 
hysteresis loop in the X-ray hardness-intensity diagram. We also find that, 
during the final stages of the HS state the source becomes softer in color and 
occupies a new region in the color-color plane. The hysteresis loop, as well 
as presence of this `transitional soft state'(TSS) are hard to explain in the 
regime of standard single flow accretion models.

In \S2 we describe our observations and data analysis techniques. The results 
of broad-band spectral color analysis is described in \S2.1. A detailed 
spectral deconvolution is presented in \S2.2. The temporal regime between 
0.1Hz - 4kHz is analyzed in \S2.3, with emphasis on the continuum structure of 
the low frequency power (\S2.3.1) and search for high frequency quasi-periodic 
oscillations (\S2.3.2). In \S3 we describe the transitional state during decline
associated with the hysteresis loop. Section 4 summarizes our results and 
speculates on possible physical mechanisms that might be causing the observed 
spectral and temporal features and their evolution.

\section{Observations and Data Analysis}

The quiescent flux from Aql X-1 is too low for the {\it All Sky Monitor} (ASM) 
\citep{l1996} on the Rossi X-ray Timing Explorer satellite (RXTE) 
\citep{brs1993} to detect. Even during the rising phase of the outburst, by 
the time the ASM starts seeing counts, the source is already quite bright in 
the Proportional Counter Array (PCA)\citep{j1996}. However, from previous 
observations it has been observed that the optical flux for Aql X-1 and other 
SXTs rises several days before the X-ray \citep{o1997,j2001}. Therefore ground 
based optical monitoring using the Yale 1m telescope at CTIO, Chile, operated 
by the YALO consortium \citep{b1999} was used to trigger the series of 
target-of-opportunity (TOO) X-ray observations. The additional lead time 
provided by our daily optical monitoring program allowed us to observe the 
source from the very early phase of the outburst. Observations continued until 
the source again faded below detection limit about two months later.

\begin{figure} 
\begin{center}
\includegraphics[height=4.0in]{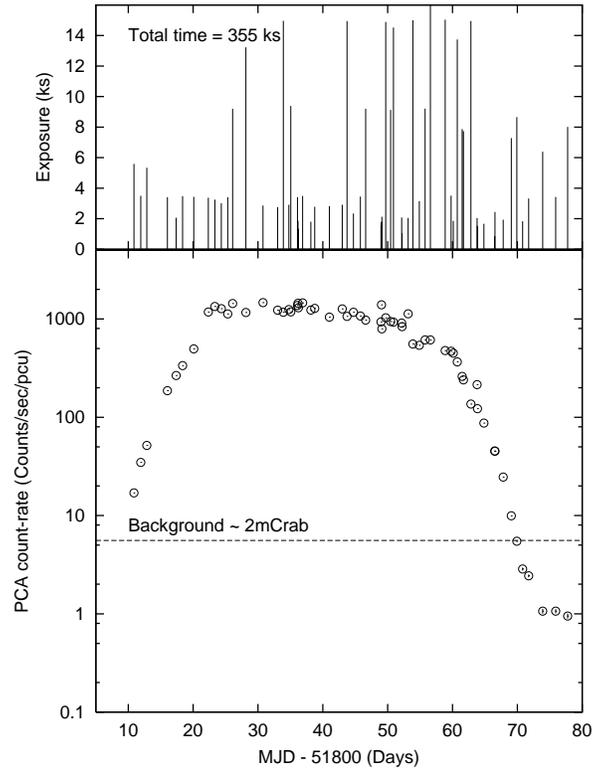} 
\caption{RXTE data from 2000 outburst of Aql X-1. (Top) Exposure times for the 
63 PCA dwells.  (Bottom) Full-range ($\approx$ 2-60 keV) background subtracted 
PCA light curve.\label{exposure}}
\end{center}
\end{figure}

The data presented in this paper were collected with the PCA instrument on 
RXTE. Our observations of Aql X-1 started from 24th September 2000 
(MJD 51811.3). We observed until the end of November 2000 when it again faded 
below RXTE's detection limit. The source was observed for a total of about 355 
ks divided in 63 pointed PCA observations. The times reported in this work are 
represented as offsets from MJD 51800. In Fig.~\ref{exposure} the distribution 
of observing time is shown on top panel and the background subtracted 
PCA ($\approx$ 2-60 keV) lightcurve at the bottom. The first four columns of 
Table~\ref{tab:obslog} gives the serial number of the observation, observation 
IDs, observation start time in MJD and exposure times in kiloseconds. Our first
 observation is at MJD$\sim 51814$ days. Observations were conducted 
typically once a day. Between MJD$\sim$ 51820 days and 51822 days, the total 
count-rate jumped from $\sim$475 counts/sec/pcu to $\sim$1175 counts/sec/pcu. 
This sudden jump was accompanied by a sharp change from ``low-hard'' spectral 
state to ``high-soft'' state (discussed in in detail later). However lack of 
data in-between prevented us from observing details of the transition from 
hard to soft state. The declining phase of the outburst however was much 
slower and the spectral state transitions are better resolved during the 
declining phase.

The {\it Standard 2} PCA data for the observations were extracted using the 
FTOOLS 5.0.2 package. Since the count-rates are small during the beginning and 
end of the outburst, we have used data from the observations when the maximum 
number PCUs were turned on. In order to avoid looking at the earth, we 
considered data only when the elevation angle was $10^{\circ}$ or more. Any 
data with (i) offset pointing more than $0^{\circ}.02$; or (ii) within 30 mins 
of SAA passage; or (iii) trapped electron contamination more than the 
prescribed value (i.e $ELECTRON2>0.1$) were rejected. Appropriate faint/bright 
background models were used when the source was fainter/brighter than 40 
counts/s/PCU.

\subsection{Fluxes and Colors} 
\begin{figure} 
\begin{center} 
\includegraphics[height=4.0in, angle=-90]{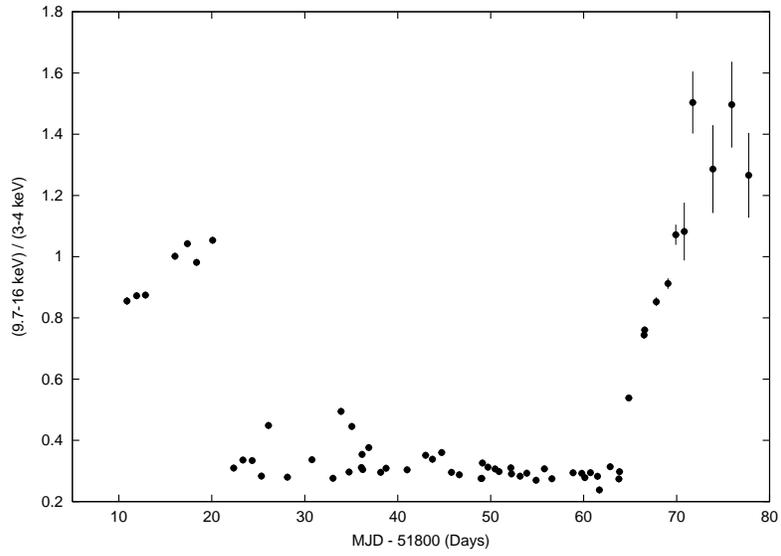} 
\caption{Variation of hardness, defined as counts in 9.7-16 keV range divided 
by counts in 3-4 keV range, with time for the 2000 outburst of Aql X-1.
\label{hardnesstime}}
\end{center}
\end{figure}

Following \citet{gd2002}, we extracted {\it standard 2} lightcurves for four 
energy bands, viz. 3-4, 4-6.4, 6.4-9.7 and 9.7-16 keV (corresponding to binned 
Std2 channels 5-6, 7-11, 13-19 and 20-33 respectively). Two bursts were 
detected during the rising phase and were excluded from analysis \citep{y2003}. 
The hardness-ratio for each observation was calculated from the ratio of counts
 in the hardest band (9.7-16 keV) to the softest band (3-4 keV; see 
Fig.~\ref{hardnesstime}). For typical spectral fits the softest band (3-4 keV) 
contains mostly thermal photons whereas the hardest band (9.7-1.6 keV) is 
dominated by powerlaw photons. Hence the (9.7-16 keV)/(3-4 keV) hardness ratio 
is an indicator of the ratio of powerlaw photons to the disk photons. As 
expected, the spectra remain hard during the low state, but make a very rapid 
transition to the soft state between MJD $\sim$ 51820 and 51822. The source 
remains soft until MJD$\sim$ 51864 after which we see a smooth transition to 
the hard state. However the transition back to the LH state from the HS state 
during the decaying phase occurs more slowly. However it is to be noted that 
since we only consider photons within certain energy-bands, the numerical 
values of the hardness-ratios do not say much about the physical processes nor 
is the hardness ratio equal to disk/powerlaw luminosity.

As seen in Fig.~\ref{xcmd}, the source traces a hysteresis loop over the 
entire outburst cycle in the hardness-luminosity plane [(9.7-16)keV/(3-4)keV
hardness-ratio vs. (6.4-9.7)keV count-rate]. Aql X-1 and few other SXT's are 
now known to show this feature \citep{mc2003,mrc2002,rmkf2000,c1998}. 
The transition from hard-to-soft state takes place at a (6.4-9.7 keV) 
luminosity $\sim10$ times higher than the luminosity at which it returns back 
to the hard state. The observed trend in evolution is very similar to what was 
observed by \citet{mc2003} using ASM data during the 1998 outburst.

A particularly interesting feature of the spectral evolution emerges when one 
considers the evolution in the color-color diagram. Soft color was defined as 
a ratio of 4-6.4 to 3-4 keV count rates, and hard color as a 9.7-16 over 
6.4-9.7 keV ratio. We plotted each observation (dwell) as one point on the 
color-color diagram. Since the evolution of the source on the color-color 
plane during an observation is small compared to the entire space it spans 
during the entire outburst cycle, this is a reasonable procedure to track the 
time evolution during the entire outburst (Fig.~\ref{colcol}, top panel). We 
find that the overall soft state (hard color $\le$0.4) has two sub-states. 
Once the source makes a transition from hard state to the soft state, it 
occupies the lower right of the CC diagram (shown by open circles in the 
bottom panel of Fig.~\ref{colcol}) and stays there as long as the luminosity 
is close to that of peak outburst (plateau in Fig.~\ref{exposure}). Once the 
decay starts appreciably, the source moves slightly to the left in CC plane 
(filled circles in Fig.~\ref{colcol}), corresponding to the TSS, and stays 
there until it makes another sharp transition to the initial hard state. The 
HS state and the TS state occupy disjoint (though neighboring) regimes of the 
color-color diagram.

\subsection{Spectral deconvolution}
Spectral analysis was carried out with the XSPEC software (v.11.0.1). We used 
the PCA {\it Standard 2} data in the energy range of 3-20 keV where PCA 
calibrations are the best. The spectral model we adopted consist of two 
additive components: a multi-colored disk (MCD) or disk blackbody (DBB) 
\citep{m1984}, plus a powerlaw. Galactic X-ray absorption was accounted for 
using the multiplicative photoelectric absorption model {\it wabs} 
\citep{mm1983}, with a constant hydrogen column density of $3.4\times10^{21}$ 
atoms/cm$^2$, the galactic value along the line of sight towards Aql X-1. It 
was found that the inclusion of another multiplicative component, a smeared 
edge ({\it smedge} in XSPEC terminology) greatly improved the spectral fits. 
The threshold energy of the edge was constrained to lie between 7-8 keV. 
Assuming a systematic error of 1\% for all PCA channels \citep{h2001,mc2003}, 
the above model gave good fits with $\chi^2/dof$ close to 1. Including an Iron 
line near 6.5 keV improved the fits only marginally and hence we have omitted 
the feature. 

The results of the fits are tabulated in Table~\ref{tab:obslog} and plotted in 
Fig.~\ref{spectral}. As expected, both the disk and powerlaw normalizations 
rise/fall during the rise/decay of the outburst. In the hard state, the powerlaw 
photon index $\Gamma$ is flatter ($<2$) compared to its value in the soft state 
when it is always steeper than 2. Remarkably, all but the disk normalization 
remains very constant throughout the soft state. The only change when the 
source moves from the HS to the TSS is that the disk normalization falls as 
shown in Fig.~\ref{fits}.

\subsection{Temporal Analysis}
We performed a Leahy-Normalized fast Fourier transform \citep{l1983} for 
every 64s segment using {\it Event} mode data of each observation for the full 
PCA ($\sim$2-60 keV) energy band. The {\it powspec} tool in XRONOS package 
was used to calculate the power density spectrum (PDS). Tha data was binned 
into $2^{-6}$s time bins, corresponding to a Nyquist frequency of 32 Hz. The 
error bars in the average power spectrum were calculated by evaluating the 
standard deviation of the average power for each frequency. These indivudual 
PDSs were then added to obtain average power-density spectrum for each dwell. 
The PDS were rebinned logarithmically to increase S/N. Finally white Poisson 
noise was subtracted and the PDS renormalized by the mean 
count rate, as discussed by \citet{vdk1995}. Thus the final PDS has 
the units of (rms/mean)$^2$ Hz$^{-1}$. The total time variability in a certain 
frequency range was estimated by fitting the corresponding PDS with a 
power-law or a combination of power-law and Lorentzian, and integrating over 
the frequency range. Most of the power comes from frequencies less than 30 Hz, 
beyond which the spectra is essentially dominated by white Poisson noise.

\subsubsection{Low Frequency Power}
During the LH rising phase, the power spectra in the range of 0.1-30 Hz exhibit
 a shallow powerlaw continuum up to a break frequency $f_{br}$. Beyond the 
break frequency the noise falls as $f^{-1}$. To describe the continuum 
behavior we fitted a broken power law model of the form:

\[{\rm PDS}(f) = \left \{ \begin{array}{r}
        C\left( \frac{f}{f_{br}} \right)^{\alpha_{1}}, f < f_{br} \\
        C\left( \frac{f}{f_{br}} \right)^{\alpha_{2}}, f \ge f_{br} \\
\end{array}
\right .
\]

where $\alpha_{1}$ and $\alpha_{2}$ are power law indices, and $f_{br}$ is
defined above.  The constant $C$ is chosen so that the integral of the
continuum model gives the correct power, expressed as fractional rms
variability.

Broad QPO-like humps were also observed in some of the spectra 
(see Fig.~\ref{pdsfig}). With the progress of the outburst the break frequency 
as well as the QPO frequency appears to increase \citep{y2003} whereas the 
total continuum power level decreases. This trend was also observed during the 
1999 outburst of the same source \citep{j2001} and for Cyg X-1 
\citep{bh1990,b2002}. Table~\ref{tab:pds} lists the break frequency, powerlaw 
indices and the relevant QPO parameters for these dwells, averaged over the 
exposure times in each dwells.

The soft state power spectrum is relatively featureless, the rms amplitudes 
are small and can be modelled by a flat powerlaw at frequencies lower than 
$\sim$ 10 Hz, and dominated by Poisson noise residuals at higher frequencies. 
Occasionally the signature of a break or a hump is observed, but the errors 
are too large to identify any feature conclusively, or to trace the evolution 
of any feature.

The variation of total power within 0.1-30 Hz frequency range (expressed in 
percentage rms variability), over the entire outburst is shown in 
Fig.~\ref{rmstime}. The strong correlation of the rms variability with 
spectral hardness (Fig.~\ref{hardnesstime}) strongly suggests that most of the 
time variability indeed comes from the hard, nonthermal photons presumably 
Compton upscattered in the corona above the accretion disk. The rms energy 
spectra for a particular dwell (Fig.~\ref{rmsE}) also supports the argument 
that most of the time variability comes from hard photons.

\subsubsection{kHz QPOs}
We also performed a Leahy-Normalized fast Fourier transform for every 128s 
segment of the {\it Event} data sets, extending up to 4 kHz in frequency. The 
frames were weighted and co-added to obtain average power-density spectrum for 
each observation. We found nine kilohertz QPOs in various segments of our 41st 
and 47th-51st observations. All the QPOs were observed during the decaying 
phase of the outburst, when the source was in the TSS. Near the kHz QPOs, the 
continuum is essentially flat with variations solely due to Poisson 
fluctuation in the data. The QPO profile is modelled by a simple Lorentzian of 
the form
\[PDS(f)=\frac{(rms_{qpo})^2}{\pi}\frac{\Delta f/2}{(f-f_{qpo})^2+(\Delta f/2)^2}\]
where $rms_{qpo}$ is the integrated fractional rms amplitude of the QPO, 
$f_{qpo}$ is the center of the Lorentzian and $\Delta f$ is its full width at 
half maximum (FWHM)\citep{c1997,c1998,z1998}. Best fit estimates for the 
observed QPOs are given in Table~\ref{tab:kHz}. In Fig.~\ref{4kpow} we show a 
sample time-averaged PDS from 0.01 Hz to 4 kHz for the dwell starting MJD 
51864.17. There is no observable change in the powerspectrum or QPO frequency 
during the dwell. The last QPO occurs just before the source makes the 
transition back to hard state from the soft state. We searched the frequency 
regime up to 4kHz but there were no evidence of any other harmonics.

Previous work on kHz QPOs has shown an interesting correlation between source 
luminosity and QPO frequency \citep{vdk2001,k1998}. It 
is observed that on a smaller timescale of hours (e.g. within a single 
observation dwell), QPO frequency is well correlated with the flux. However 
this correlation is lost in larger timescales of days (e.g between 
observations). This leads to formation of ``parallel tracks'' in 
$\nu_{QPO}-Flux$ plane. This has been related to existence of two accretion 
flows rather than one. Owing to comparatively large number of kHz QPOs 
observed during the 2000 outburst of Aql X-1 and relatively high photon count 
rate, we could carry out FFT of the data for every 128s segment and track the 
QPO evolution (Fig~\ref{parallel}). We find that at lower count rates the 
tracks are somewhat more correlated and show the ``parallel tracks'' more 
clearly. At higher count-rates, however, even the short-term correlation seems
 to break down. This is in agreement with the observations of kHz QPOs in 
4U 1608-52 by \citet{m1999} (see their Fig.2).

\section{The Transitional Soft State}
As seen in \S2.1.2 and Fig.~\ref{xcmd}, once the source goes into the soft 
state, it stays in the bottom right corner of the color-luminosity diagram 
(shown by the filled triangles). It stays in this bottom right corner while 
the X-ray luminosity remains close to the peak ouburst value. In the 
color-color diagram also the soft state stands out as a separate region, 
occupying the right side of Fig.~\ref{colcol}. We associate this state with the 
canonical soft state of Aql X-1.

However, as the luminosity decreases from the peak luminosity (the region 
labelled as ``Trans Soft'' in Fig.~\ref{spectral}), the source still stayed in 
the soft state for about $\sim 10$ days before going back to the hard state, 
thereby tracing the hysteresis loop. During this time when the total source 
luminosity was below that of the state transition during the rise the source 
displayed small but consistent differences from the canonical soft state. 
Thus as the source traces the hysteresis loop, it appears to enter a 
different state (or substate) that has no counterpart during the rising phase. 
All the kHz QPOs were observed when the source was in this transitional soft 
state only.

In the color-color plane, the data taken during the decline (but still in 
overall soft state) also occupies a region significantly different from the 
others. When the source is in the canonical soft state, the soft color has a 
value $>1.87$, whereas it is softer than this value during the transitional 
soft state. In contrast, the hard color remains at almost similar values 
during both substates of the soft state. In the spectral deconvolution 
(Fig.~\ref{spectral}), the signature of this new state is observed in the 
decrease of disk normalization. Paradoxically, the decrease in the disk flux 
results in a softer spectrum in the RXTE band passes, since the powerlaw, which
 does not change, contributes a larger fraction of the flux in the soft band 
than the medium bands (see Fig.~\ref{fits}). All other spectral parameters 
viz. the spectral shapes of the disk and powerlaw as well as the powerlaw 
strength, remain the same over both the HS and TSS. 

We suggest that this ``transitional soft state'' should be considered 
separately from the canonical soft state. The presence of this state 
during decline, but not during the rise, provides additional evidence for the existence of a second parameter, a physical parameter other than 
$\dot M$, that governs state transition in STXs.

\section{Discussion}
In the context of two flow models of accretion discussed by Smith, Heindl, \& 
Swank (2002, hereafter SHS), accretion can take place either via the accretion 
disk or through a hot sub-Keplerian halo. The disk photons are soft and have a 
thermal origin whereas the halo photons comprise the hard powerlaw photons. 
Any change in the disk mass accretion rate occurs only on viscous timescales 
which are much longer than the almost free-fall timescales in the halo. Close 
to peak outburst luminosities, the disk photons dominate the spectrum and 
according to SHS one should see the {\em static} soft state. A {\em dynamic} 
soft state is seen when the overall mass inflow rate starts decreasing, to 
which the halo responds much faster than the disk and hence drains out its 
contents thereby making the spectrum softer. If the two soft states that we 
observed are indeed the static and dynamic soft states proposed by SHS, then 
the $\sim10$ day decay of Aql X-1 in the dynamic soft state is a direct 
measurement of viscous timescale of the accretion disk. However, the spectral 
fits to the data show that the softening in soft color [(4-6.4)keV/(3-4)kev 
countrates] during the transitional soft state is in fact due to a decrease of 
flux in the disk, not in the powerlaw. This is contrary to what we would 
expect from the picture put forward by SHS where once the overall $\dot M$ 
starts declining, the corona responds faster and drains out sooner than
compared to the disk. Thus the SHS model predicts that the powerlaw 
normalization will fall rather than the disk. Therefore while our data 
disagree with the particular mechanism discussed by SHS, they provide strong 
confirmation of the general idea that $\ddot M$, as well as $\dot M$, is an 
important factor in governing state transitions. 

The now well established fact that the hard to soft state transition occurs at 
a higher luminosity than that of soft to hard transition rules out one flow 
propeller models of accretion for Aql X-1. As discussed by \citet{mc2003}, the 
strong ADAF is also not tenable. The occurrence of this hysteresis phenomenon 
for all the recent outbursts of Aql X-1 further demonstrates that mass 
accretion rate is not the only factor in state transition in NS/BH 
systems.

From a compilation of RXTE data of several transient atoll sources, 
\citet{gd2002} and \citet{mrc2002} have shown that the atolls form a ``Z'' 
shaped track in the color-color diagram, similar to Z sources. They conclude 
that both atolls and Z sources move along their tracks on the color-color 
plane and do not jump between the tracks. For the source 4U 1705-44 
\citet{gd2002} analysed the source movement with time and confirmed that the 
motion in the diagram goes along the track and they did not find any jump 
between the branches. Excellent coverage of the 2000 outburst of Aql X-1 
allowed us to track its time evolution on the color-color plane quite closely. 
We found that although the overall time evolution does form a Z shape, the 
source jumps from the middle branch to the lower right instead of smoothly 
following a Z (Fig.~\ref{colcol}). Thus the usual curve length parameter S 
does not describe the time evolution of the source for the outburst described 
here.

\begin{figure}
\begin{center} 
\includegraphics[height=4.0in, angle=-90]{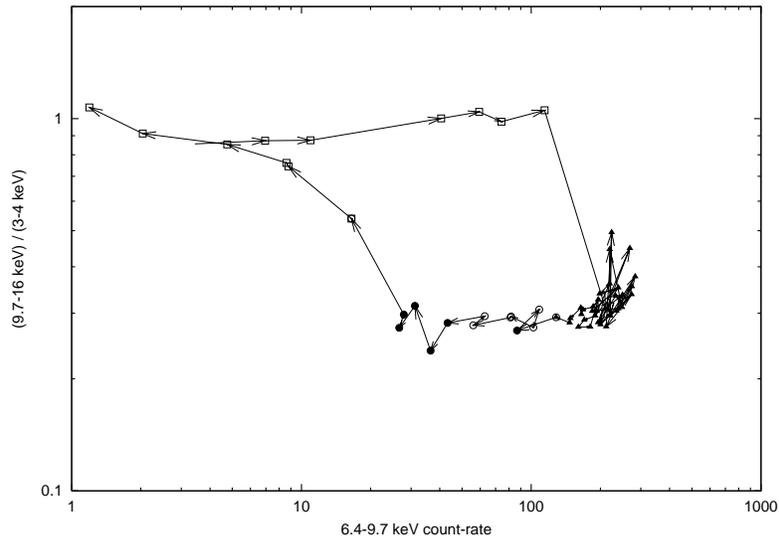}
\caption{Evolution of Aql X-1 on X-ray color luminosity diagram. The arrows 
point in the direction of advancing time. Observations corresponding to the 
source being in the {\it hard state} are shown as open squares, the {\it 
canonical soft state} is shown by filled triangles and {\it transitional soft 
state} is shown by circles. The errorbars become larger towards the left 
(because very few photons are detected during the very early as well as very 
late phase of the outburst). However for the range of hardness and intensity 
plotted, the errorbars are smaller than the symbol size and hence not plotted. 
The dwells for which kHz QPOs were detected are indicated by filled circles.
\label{xcmd}}
\end{center}
\end{figure}

\begin{figure}
\begin{center} 
\includegraphics[height=7.5in, angle=0]{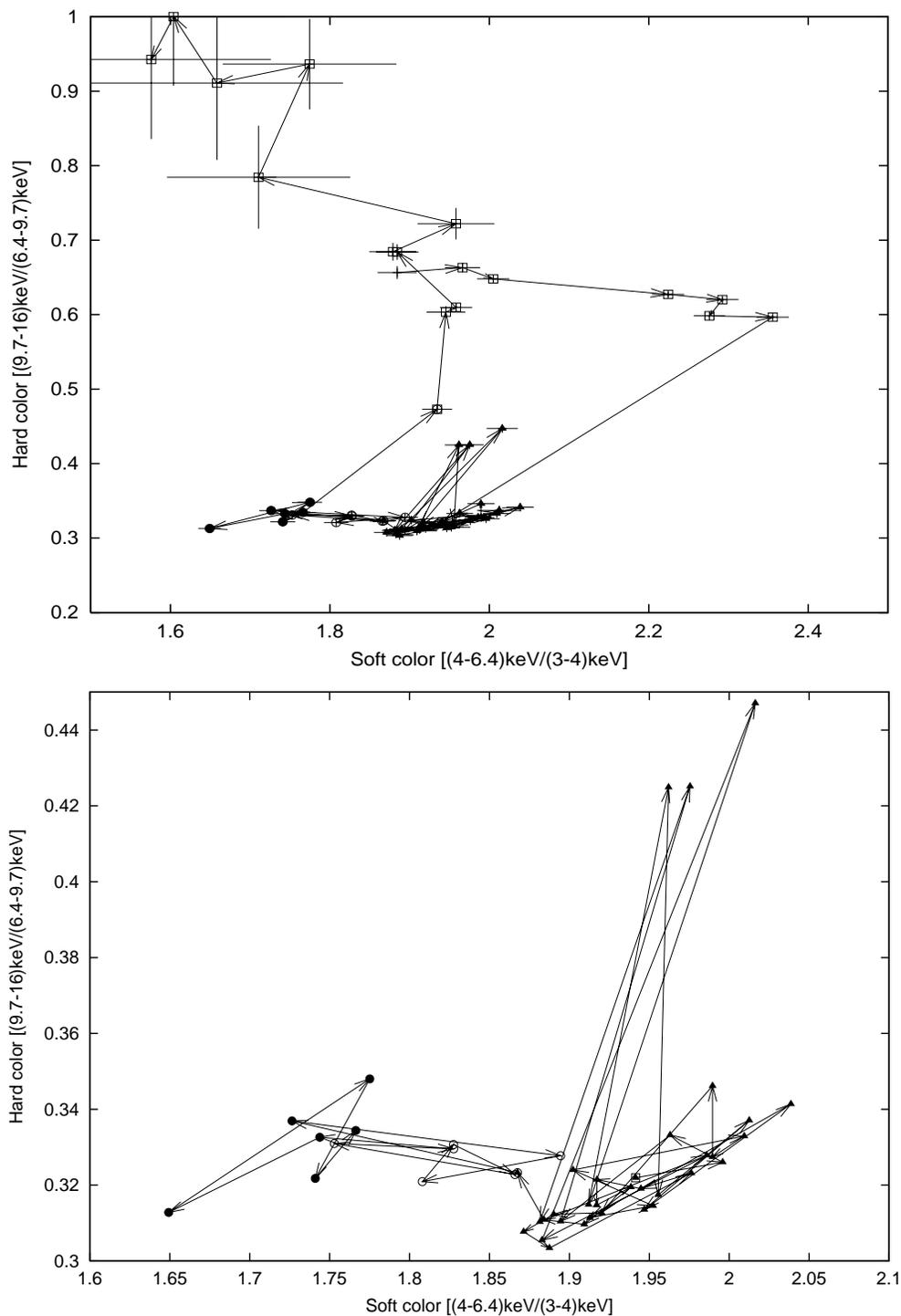} 
\caption{(Top) Evolution on the Color-color plane during the entire outburst.
The symbols have the same meaning as in Fig.~\ref{xcmd}. (Bottom) Blow-up of 
the region corresponding to the soft state. The errorbars have not been 
plotted in the bottom plot to avoid confusion.\label{colcol}}
\end{center}
\end{figure}

\begin{figure}
\centering
\includegraphics[height=7.5in, angle=0]{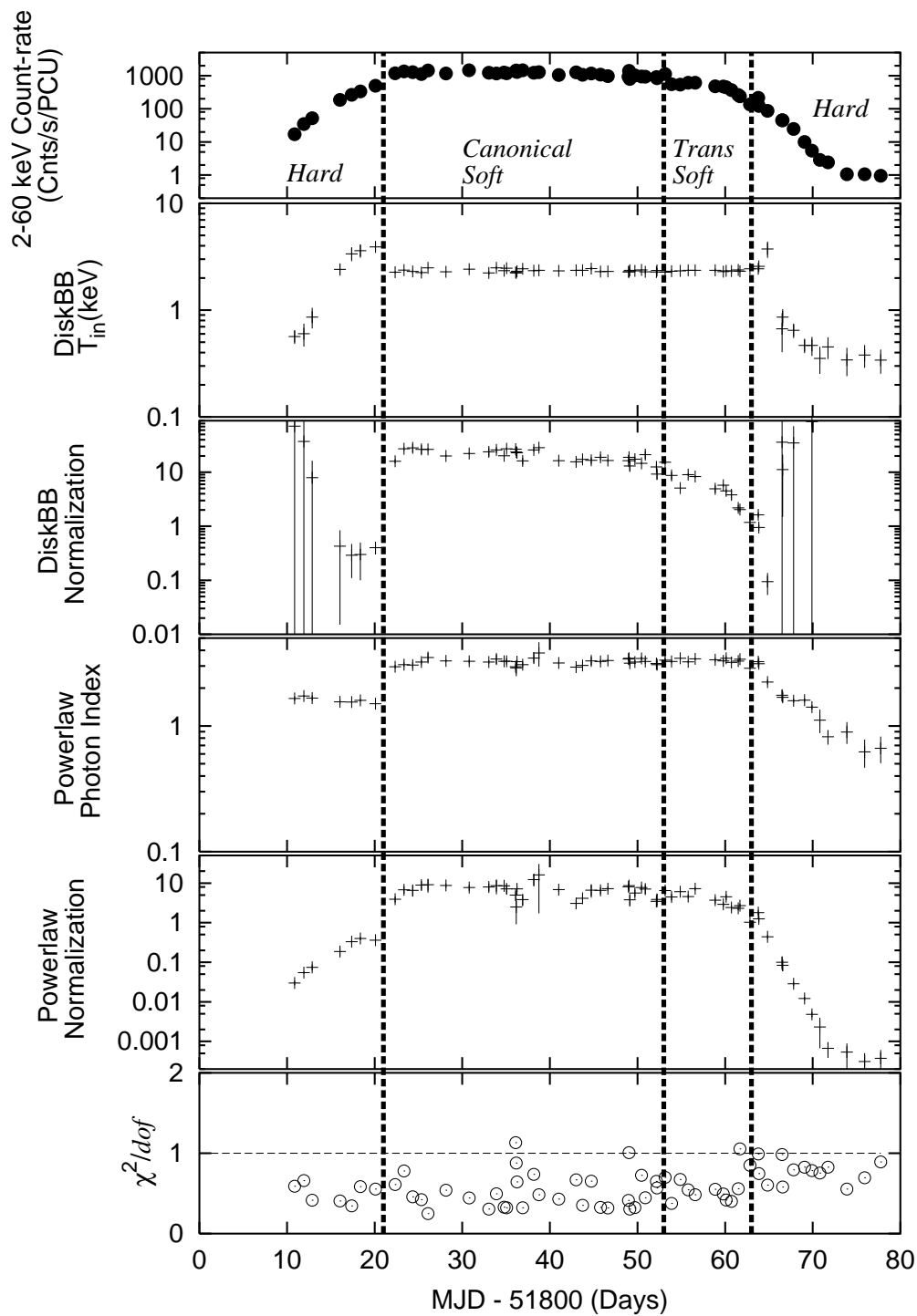}
\caption{Spectral deconvolution, evolution of the spectral fit parameters with 
time during the outburst of Aql X-1. The different spectral states as defined 
from the color-color plot (Fig~\ref{colcol}) are separated by the thick dashed 
lines.\label{spectral}}
\end{figure}

\begin{figure}
\centering
\includegraphics[width=7.0in, angle=0]{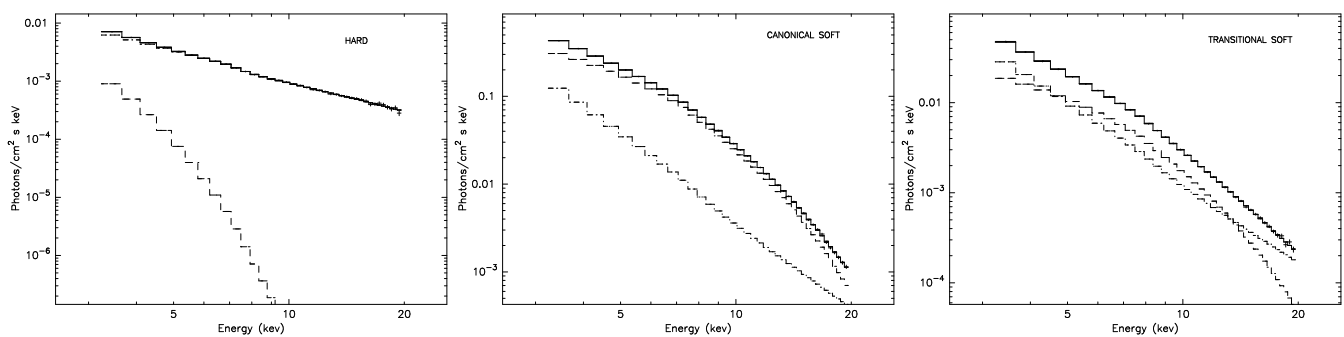}
\caption{Sample unfolded spectra during the various states. Left: Hard state 
spectrum for the dwell on MJD 51812.31. Middle: Canonical soft state as seen 
on MJD 51836.63. Right: Transitional soft state on MJD 51863.24. The dashed 
histogram is the fitted disk blackbody, dash-dotted histogram is the powerlaw, 
solid histogram is the overall fit and the crosses (only visible near higher 
end of the spectra) are the data points. Note that the primary difference 
between the canonical soft state and the transitional soft state is the lower 
relative normalization of the disk component in the transitional soft state.
\label{fits}}
\end{figure}

\begin{figure}
\includegraphics[height=7.5in, angle=0]{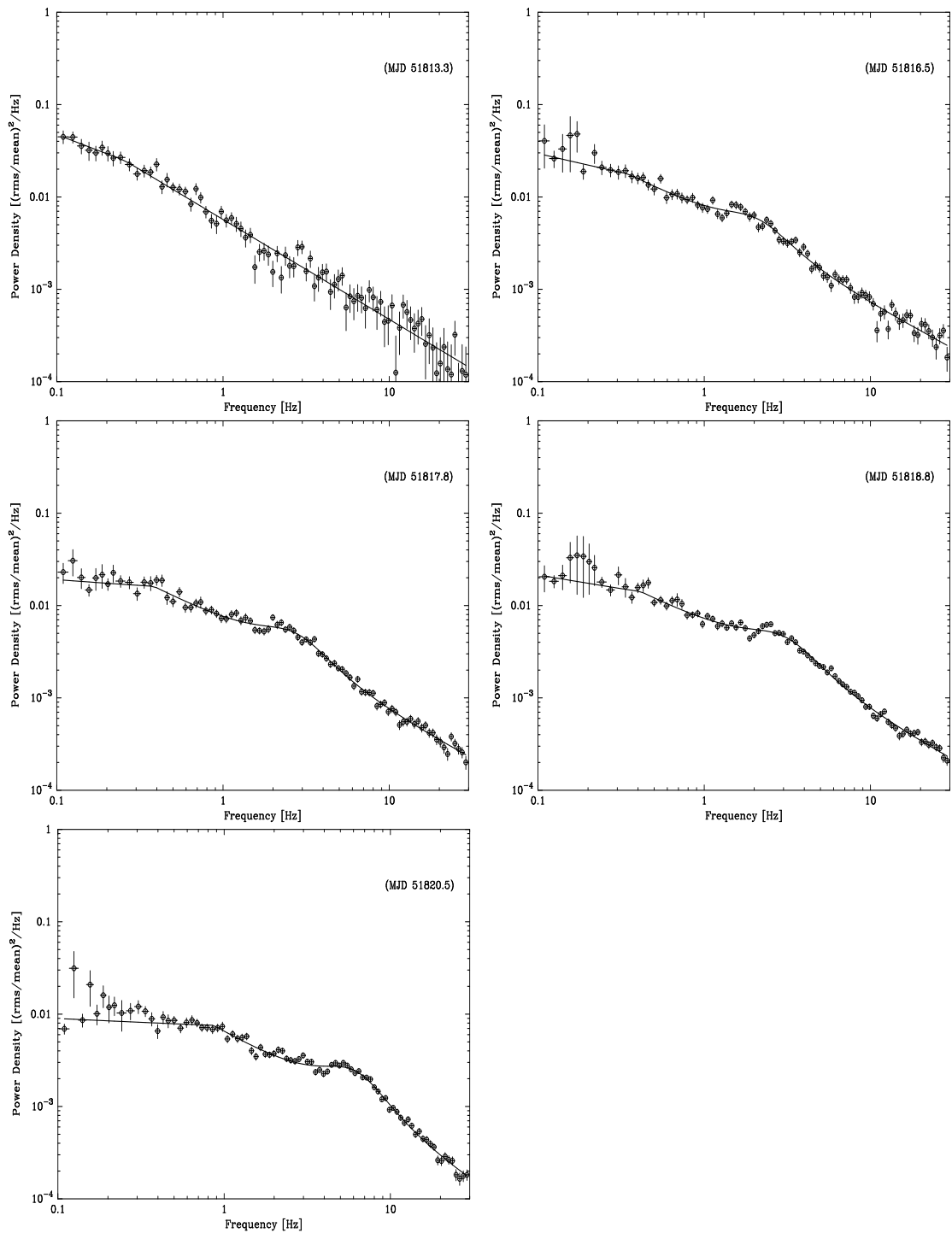}
\caption{Low frequency PDS during the rising phase. The time of observation is 
shown in top right corner for each plot. The ordinate represents the Poisson 
noise subtracted PDS renormalized by the mean count rate, plotted against 
temporal frequency along the abscissae.\label{pdsfig}}
\end{figure}

\begin{figure}
\centering
\includegraphics[height=4.0in, angle=-90]{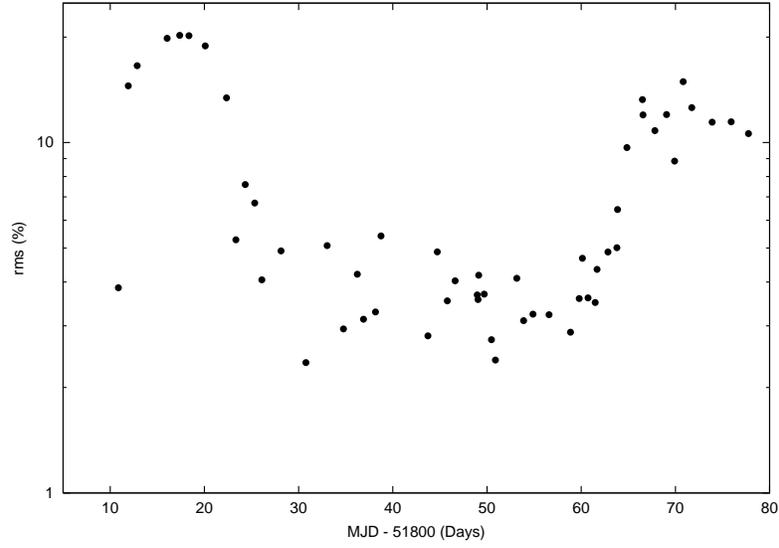}
\caption{Evolution of time-variability. The total rms variability in the 
Fourier power spectrum between 0.1-30 Hz is plotted against time. Each dwell 
is one point in the diagram.\label{rmstime}}
\end{figure}

\begin{figure}
\centering
\includegraphics[height=4.0in, angle=-90]{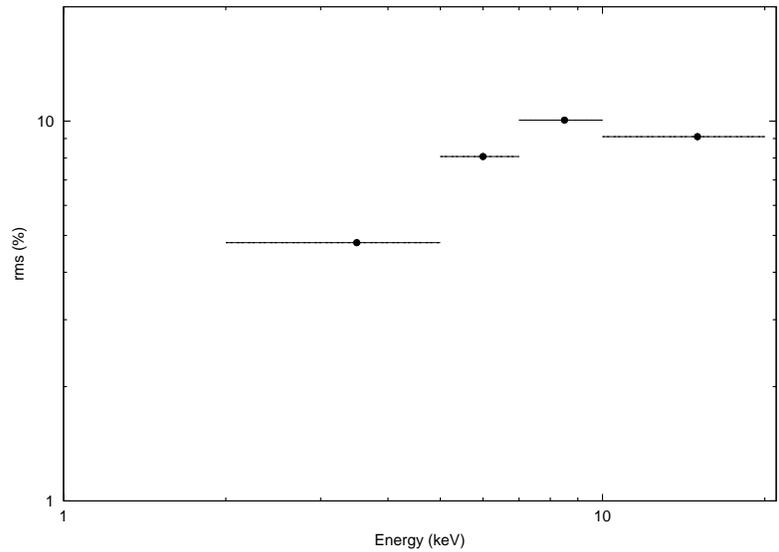}
\caption{Rms energy spectra for a soft state dwell (t=51863.78 day). 
Percentage rms variability for the energy ranges 2-5, 5-7, 7-10 and 
10-20 keV are plotted.\label{rmsE}}
\end{figure}

\begin{figure}
\centering
\includegraphics[width=4.0in, angle=0]{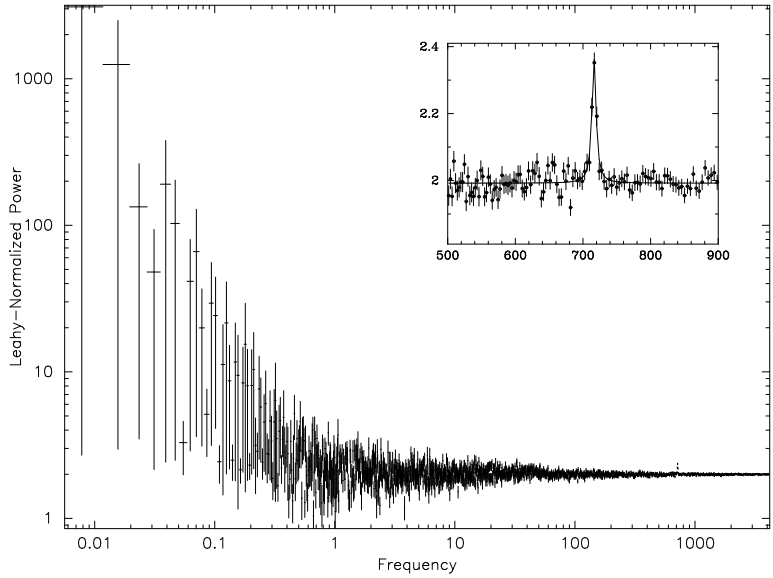}
\caption{A sample time-averaged PDS from 0.01 Hz to 4 kHz for the dwell 
starting MJD 51864.17. The region around the QPO is blown up and shown in the 
inset. The spectrum above 100 Hz is essentially dominated by Poisson counting 
statistics.
\label{4kpow}} 
\end{figure}

Interestingly, all the kHz QPOs seen in the data for this outburst were 
observed when the source was in the transitional soft state, in particular, 
once the 6.4-9.7 keV flux falls below $\sim50$ counts/sec/PCU 
(Fig.~\ref{xcmd}). The kHz QPOs observed by \citet{z1998} during the 1997 
outburst of Aql X-1 also were seen during the declining phase, when the 
luminosity was considerably smaller than the peak value. In contrast 
\citet{c1998} reported kHz QPOs during the rising phase of the outburst of 
1998. This might suggest that the TSS does occur in the rise, but has been 
missed because the rise is so much faster than the decline. But in all these 3 
outbursts, kHz QPOs were observed during intermediate luminosity and never 
during very high or very low luminosity. Even the high frequency QPOs observed 
by \citet{rem2002} for black hole systems occur when the source is in a 
transitional state where the spectra is neither entirely dominated by the soft 
photons nor dominated by entirely hard photons.

\begin{figure}
\begin{center}
\includegraphics[height=4.0in, angle=-90]{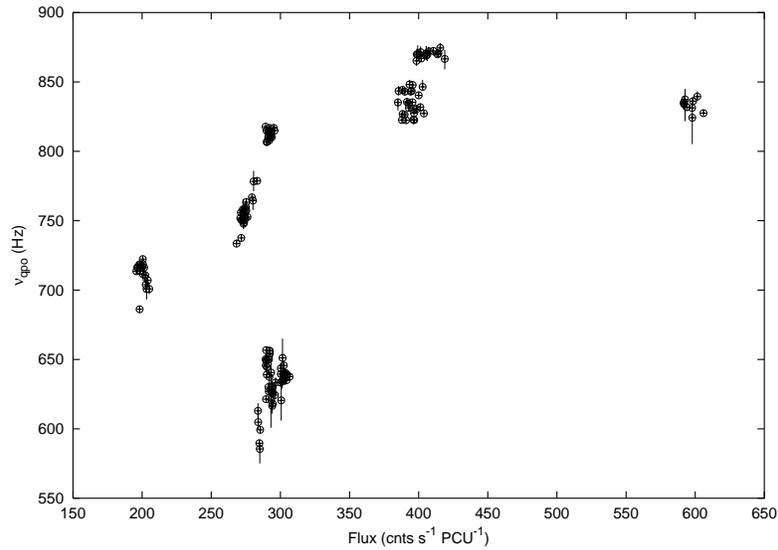}
\caption{Observed kHz QPO frequency vs. count rate showing the ``parallel 
tracks'' as they occur during the 2000 outburst of Aql X-1.\label{parallel}}
\end{center}
\end{figure}

If these QPOs are blobs of matter in Keplerian orbit around the compact object,
then their radial distance (in km, given by $r_{km}$)from the compact star is  
$r_{km}\approx15(\frac{m}{\nu_{qpo;kHz}})^{2/3}$, where $m$ is the mass of the 
accretor in solar units and $\nu_{qpo;kHz}$ is the observed QPO frequency in 
kiloHertz. In this model the highest QPO frequency during the outburst 
observed around $\sim870$ Hz puts the minimum radius of the QPOs between 
17.5 and 23.7 km depending on the neutron star mass (assumed to be between 
$1.2M_{\odot}$ and $3M_{\odot}$). According to the Keplerian orbit model of 
QPO origin, an increase in inferred mass accretion rate pushes the inner edge 
of the accretion disk inwards and therefore predicts QPO frequency to increase 
with $\dot M$ and saturate beyond certain $\dot M$. In Fig.~\ref{parallel} we 
may see this saturation beyond $\sim400$ counts/sec/PCU. However the fact 
that there are no QPOs when the source is either very faint or very bright 
supports the hypothesis put forward by \citet{c2000} that the root of these 
kHz QPO lie in disk-magnetosphere interaction. The existence of a higher 
luminosity barrier can be regarded as an evidence for the presence of an 
innermost stable orbit.

DM would like to thank Tom Maccarone and Aya Kubota for their help with 
software. It is a pleasure to ackonwledge the help of Paolo Coppi for 
many informative discussions and useful comments.
This work was supported by National Science Foundation grant AST 00-98421 and 
NASA ADP grant NAG5-13336.

\clearpage

\clearpage
%% Table 1.
\begin{deluxetable}{lllllll}
\tabletypesize{\footnotesize}
\tablecolumns{6}
\tablewidth{0pc}
\tablecaption{Results of Spectral Deconvolution \label{tab:obslog}}
\tablehead{
\colhead{Srl. No.}&\colhead{ObsID}&\colhead{Time (MJD)\tablenotemark{\star}}
&\colhead{Exposure (ks)\tablenotemark{\dag}}&\colhead{$T_{in}$(keV)}
&\colhead{Photon Index}&\colhead{$(\chi^2/dof)$\tablenotemark{\ddag}}}
\startdata
1 & 50049-01-03-00 & 51811.26 & 5.59 & 0.57$\pm$0.08 & 1.65$\pm$0.06 & 0.59 \\
2 & 50049-01-03-01 & 51812.31 & 3.50 & 0.60$\pm$0.15 & 1.73$\pm$0.03 & 0.66 \\
3 & 50049-01-03-02 & 51813.25 & 5.34 & 0.86$\pm$0.19 & 1.67$\pm$0.08 & 0.41 \\
4 & 50049-01-04-00 & 51816.43 & 3.41 & 2.40$\pm$0.30 & 1.56$\pm$0.07 & 0.41 \\
5 & 50049-01-04-01 & 51817.77 & 2.06 & 3.37$\pm$0.50 & 1.55$\pm$0.04 & 0.35 \\
6 & 50049-01-04-02 & 51818.76 & 3.48 & 3.60$\pm$0.50 & 1.60$\pm$0.04 & 0.58 \\
7 & 50049-01-04-03 & 51820.48 & 3.43 & 3.90$\pm$0.40 & 1.50$\pm$0.06 & 0.55 \\
8 & 50049-01-04-04 & 51822.74 & 3.37 & 2.26$\pm$0.05 & 2.96$\pm$0.10 & 0.61 \\
9 & 50049-01-05-00 & 51823.73 & 3.25 & 2.37$\pm$0.04 & 3.07$\pm$0.13 & 0.78 \\
10 & 50049-01-05-01 & 51824.73 & 3.01 & 2.30$\pm$0.09 & 3.05$\pm$0.17 & 0.46 \\
11 & 50049-01-05-02 & 51825.73 & 3.41 & 2.25$\pm$0.09 & 3.22$\pm$0.14 & 0.42 \\
12 & 50049-02-01-00 & 51826.49 & 9.21 & 2.49$\pm$0.04 & 3.50$\pm$0.30 & 0.25 \\
13 & 50049-02-02-00 & 51828.53 & 13.23 & 2.29$\pm$0.03 & 3.29$\pm$0.08 & 0.54 \\
14 & 50049-02-03-01 & 51831.17 & 2.86 & 2.42$\pm$0.03 & 3.26$\pm$0.11 & 0.44 \\
15 & 50049-02-03-00 & 51833.42 & 2.77 & 2.23$\pm$0.08 & 3.23$\pm$0.10 & 0.31 \\
16 & 50049-02-04-00 & 51834.29 & 14.97 & 2.49$\pm$0.08 & 3.40$\pm$0.30 & 0.50 \\
17 & 50049-02-03-02 & 51835.15 & 2.91 & 2.34$\pm$0.03 & 3.27$\pm$0.08 & 0.33 \\
18 & 50049-02-05-00 & 51835.45 & 9.39 & 2.47$\pm$0.03 & 3.27$\pm$0.18 & 0.32 \\
19 & 50049-02-06-00 & 51836.48 & 3.42 & 2.25$\pm$0.06 & 2.96$\pm$0.17 & 1.13 \\
20 & 50049-02-06-01 & 51836.56 & 1.87 & 2.23$\pm$0.09 & 2.90$\pm$0.42 & 0.88 \\
21 & 50049-02-06-02 & 51836.63 & 1.34 & 2.31$\pm$0.03 & 3.25$\pm$0.11 & 0.64 \\
22 & 50049-02-07-00 & 51837.28 & 3.50 & 2.44$\pm$0.08 & 3.07$\pm$0.17 & 0.32 \\
23 & 50049-02-07-01 & 51838.55 & 1.80 & 2.34$\pm$0.05 & 3.46$\pm$0.28 & 0.74 \\
24 & 50049-02-07-02 & 51839.13 & 2.78 & 2.36$\pm$0.06 & 3.81$\pm$0.83 & 0.48 \\
25 & 50049-02-07-03 & 51841.40 & 2.82 & 2.33$\pm$0.03 & 3.17$\pm$0.08 & 0.43 \\
26 & 50049-02-07-04 & 51843.39 & 2.91 & 2.36$\pm$0.06 & 2.94$\pm$0.13 & 0.67 \\
27 & 50049-02-08-00 & 51844.12 & 14.94 & 2.36$\pm$0.08 & 3.02$\pm$0.11 & 0.35 \\
28 & 50049-02-08-01 & 51845.11 & 2.35 & 2.45$\pm$0.09 & 3.29$\pm$0.26 & 0.65 \\
29 & 50049-02-08-03 & 51846.18 & 3.46 & 2.30$\pm$0.06 & 3.25$\pm$0.12 & 0.33 \\
30 & 50049-02-09-00G & 51847.01 & 9.20 & 2.31$\pm$0.03 & 3.32$\pm$0.08 & 0.32 \\
31 & 50049-02-10-03 & 51849.36 & 1.57 & 2.28$\pm$0.03 & 3.41$\pm$0.11 & 0.41 \\
32 & 50049-02-10-02 & 51849.43 & 1.79 & 2.28$\pm$0.03 & 3.46$\pm$0.12 & 1.01 \\
33 & 50049-02-10-01 & 51849.50 & 2.13 & 2.35$\pm$0.10 & 3.16$\pm$0.16 & 0.30 \\
34 & 50049-02-10-00 & 51850.09 & 14.89 & 2.33$\pm$0.06 & 3.22$\pm$0.11 & 0.32 \\
35 & 50049-02-10-05 & 51850.86 & 9.13 & 2.37$\pm$0.03 & 3.45$\pm$0.11 & 0.73 \\
36 & 50049-02-11-00 & 51851.28 & 14.53 & 2.28$\pm$0.09 & 3.26$\pm$0.14 & 0.45 \\
37 & 50049-02-11-01 & 51852.56 & 2.08 & 2.27$\pm$0.05 & 3.13$\pm$0.19 & 0.65 \\
38 & 50049-02-11-02 & 51852.62 & 1.02 & 2.35$\pm$0.13 & 3.07$\pm$0.28 & 0.57 \\
39 & 50049-02-12-01 & 51853.55 & 2.06 & 2.27$\pm$0.05 & 3.33$\pm$0.16 & 0.70 \\
40 & 50049-02-12-00 & 51854.27 & 15.00 & 2.31$\pm$0.09 & 3.26$\pm$0.10 & 0.37 \\
41 & 50049-02-13-00 & 51855.27 & 3.16 & 2.33$\pm$0.08 & 3.46$\pm$0.12 & 0.67 \\
42 & 50049-02-13-01 & 51856.17 & 9.21 & 2.36$\pm$0.03 & 3.23$\pm$0.08 & 0.55 \\
43 & 50049-02-14-00 & 51856.97 & 18.79 & 2.36$\pm$0.03 & 3.42$\pm$0.07 & 0.48 \\
44 & 50049-02-15-00 & 51859.25 & 15.04 & 2.36$\pm$0.09 & 3.36$\pm$0.13 & 0.55 \\
45 & 50049-02-15-01 & 51860.18 & 3.51 & 2.29$\pm$0.04 & 3.29$\pm$0.10 & 0.49 \\
46 & 50049-02-15-08 & 51860.52 & 1.86 & 2.33$\pm$0.05 & 3.45$\pm$0.14 & 0.42 \\
47 & 50049-02-15-02 & 51861.12 & 13.74 & 2.32$\pm$0.10 & 3.20$\pm$0.08 & 0.40 \\
48 & 50049-02-15-03 & 51861.88 & 7.87 & 2.38$\pm$0.07 & 3.28$\pm$0.09 & 0.56 \\
49 & 50049-02-15-04 & 51862.08 & 7.74 & 2.31$\pm$0.06 & 3.40$\pm$0.08 & 1.06 \\
50 & 50049-02-15-05 & 51863.24 & 14.96 & 2.43$\pm$0.09 & 2.87$\pm$0.06 & 0.85 \\
51 & 50049-02-15-06 & 51864.17 & 2.05 & 2.44$\pm$0.08 & 3.25$\pm$0.12 & 0.99 \\
52 & 50049-02-15-07 & 51864.24 & 1.52 & 2.57$\pm$0.10 & 3.13$\pm$0.14 & 0.75 \\
53 & 50049-03-01-00 & 51865.24 & 1.67 & 3.74$\pm$0.64 & 2.24$\pm$0.10 & 0.60 \\
54 & 50049-03-02-01 & 51866.89 & 0.85 & 0.67$\pm$0.27 & 1.75$\pm$0.19 & 0.98 \\
55 & 50049-03-02-00 & 51866.95 & 2.44 & 0.86$\pm$0.16 & 1.69$\pm$0.15 & 0.58 \\
56 & 50049-03-03-00 & 51868.20 & 1.93 & 0.65$\pm$0.10 & 1.59$\pm$0.06 & 0.79 \\
57 & 50049-03-04-00 & 51869.45 & 7.29 & 0.47$\pm$0.07 & 1.60$\pm$0.05 & 0.83 \\
58 & 50049-03-05-00 & 51870.30 & 8.66 & 0.47$\pm$0.09 & 1.41$\pm$0.06 & 0.78 \\
59 & 50049-03-06-00 & 51871.21 & 1.83 & 0.35$\pm$0.10 & 1.11$\pm$0.24 & 0.75 \\
60 & 50049-03-07-00 & 51872.13 & 3.33 & 0.45$\pm$0.10 & 0.82$\pm$0.11 & 0.83 \\
61 & 50049-03-08-00 & 51874.29 & 6.39 & 0.34$\pm$0.10 & 0.90$\pm$0.18 & 0.55 \\
62 & 50049-03-09-00 & 51876.31 & 3.43 & 0.38$\pm$0.09 & 0.62$\pm$0.16 & 0.70 \\
63 & 50049-03-10-00 & 51878.14 & 8.02 & 0.34$\pm$0.09 & 0.67$\pm$0.16 & 0.89 \\
\enddata

\tablenotetext{\star}{Observation starting time}
\tablenotetext{\dag}{Total exposure time before screening for good time intervals}
\tablenotetext{\ddag}{$dof=31$}
\end{deluxetable}

%% Table 2.
\begin{deluxetable}{lccccccc}
\tablecolumns{8}
\tablecaption{PDS Parameters for hard state dwells in rising phase \label{tab:pds}} 
\tablewidth{0pc}
\tablehead{ 
\colhead{Time(MJD)} & $f_{br}$(Hz) & $\alpha_1$ & $\alpha_2$ & $f_{qpo}$(Hz) &
FWHM(Hz) & $rms_{qpo}(\%)$}
\startdata 
51813.3 & $0.26^{+0.21}_{-0.08}$ & $-0.7^{+0.2}_{-0.2}$ & $-1.08^{+0.04}_{-0.06}$  & ... & ... & ... \\
51816.5 & $0.34^{+0.22}_{-0.11}$ & $-0.4^{+0.6}_{-0.3}$ & $-0.96^{+0.03}_{-0.03}$  & $1.8^{+0.2}_{-0.2}$ & $2.4^{+0.5}_{-0.4}$ & $10.6^{+2.0}_{-1.8}$  \\
51817.8 & $0.38^{+0.06}_{-0.06}$ & $-0.1^{+0.3}_{-0.2}$ & $-0.96^{+0.02}_{-0.02}$  & $2.3^{+0.2}_{-0.2}$ & $3.2^{+0.5}_{-0.4}$ & $12.1^{+1.5}_{-1.4}$  \\
51818.8 & $0.42^{+0.08}_{-0.12}$ & $-0.3^{+0.2}_{-0.2}$ & $-0.96^{+0.02}_{-0.02}$  & $2.5^{+0.2}_{-0.2}$ & $3.8^{+0.4}_{-0.4}$ & $13.0^{+1.1}_{-1.2}$  \\
51820.5 & $0.86^{+0.08}_{-0.12}$ & $-0.1^{+0.1}_{-0.1}$ & $-1.10^{+0.04}_{-0.04}$  & $5.4^{+0.2}_{-0.3}$ & $6.4^{+0.8}_{-0.6}$ & $13.1^{+1.2}_{-1.0}$  \\
\enddata
\end{deluxetable}

%% Table 3.
\begin{deluxetable}{lccc}
\tablecolumns{5}
\tablewidth{0pc}
\tablecaption{Observed kHz QPOs \label{tab:kHz}}
\tablehead{
\colhead{Time(Days)}&\colhead{Frequency (Hz)}&\colhead{FWHM (Hz)} &\colhead{Fractional RMS (\%)}}
\startdata
51855.29 & $833.7_{-1.2}^{+1.3} $ & $12.8_{-2.9}^{+3.9} $ & $4.4_{-0.9}^{+1.2} $ \\
51861.87 & $869.9_{-0.4}^{+0.5} $ & $07.0_{-2.1}^{+2.0} $ & $5.5_{-1.3}^{+1.7} $ \\
51861.92 & $832.5_{-1.3}^{+1.5} $ & $19.4_{-2.9}^{+3.6} $ & $6.8_{-0.9}^{+1.0} $ \\
51862.07 & $812.2_{-1.0}^{+1.2} $ & $09.8_{-1.4}^{+2.0} $ & $6.6_{-0.8}^{+1.2} $ \\
51862.12 & $754.6_{-1.1}^{+1.2} $ & $13.4_{-2.7}^{+3.2} $ & $6.6_{-1.3}^{+1.1} $ \\
51863.19 & $636.5_{-1.5}^{+1.3} $ & $16.2_{-4.1}^{+4.9} $ & $6.3_{-1.3}^{+1.6} $ \\
51863.25 & $646.2_{-2.2}^{+2.1} $ & $22.0_{-4.5}^{+6.1} $ & $6.6_{-1.3}^{+1.2} $ \\
51864.20 & $716.3_{-0.5}^{+0.6} $ & $07.3_{-1.2}^{+1.3} $ & $7.5_{-1.1}^{+1.1} $ \\
51864.27 & $702.3_{-2.1}^{+2.5} $ & $09.6_{-3.9}^{+5.1} $ & $5.8_{-2.3}^{+2.6} $ \\
\enddata
\end{deluxetable}

\newpage

%% Fig 1

%% Fig 2

%% Fig 3

%% Fig 04

%% Fig 05 

%% Fig 06 

%% Fig 07

%% Fig 08 

%% Fig 09 

%% Fig 10

%%Fig 11

\end{document}